\documentclass[preprint]{aastex} 
\usepackage{psfig} 

\slugcomment{KSUPT-02/2 \hspace{0.5truecm} July 2002}
\newcommand{\omegal}{\Omega_{\Lambda}}

\begin{document}
\title{Radio Galaxy Redshift-Angular Size Data Constraints on Dark Energy}
\author{Silviu Podariu\altaffilmark{1,2}, Ruth A. Daly\altaffilmark{3}, 
Matthew P. Mory\altaffilmark{3}, and Bharat Ratra\altaffilmark{1}}
\altaffiltext{1}{Department of Physics, Kansas State University, 
116 Cardwell Hall, Manhattan, KS 66506.} 
\altaffiltext{2}{Current address: Department of Mechanical Engineering, 
Northwestern University,
2145 Sheridan Road, B224, Evanston, IL 60208} 
\altaffiltext{3}{Department of Physics, Berks-Lehigh Valley College,
Pennsylvania State University, 
Reading, PA 19610.}
\begin{abstract} 
We use FRIIb radio galaxy redshift-angular size data to constrain 
cosmological parameters in a dark energy scalar field model. The derived 
constraints are consistent with but weaker than those determined using 
Type Ia supernova redshift-magnitude data. 
\end{abstract}
\keywords{cosmology: cosmological parameters---cosmology: 
observation---large-scale structure of the universe}
\section{Introduction} 

The last half-a-dozen years have seen a remarkable increase in the quality 
of some cosmological data. No less remarkable, but perhaps less heralded, 
has been the continuing acquisition of new types of data. These have been 
very useful developments in the on-going process of determining, through 
the cosmological tests, how well current cosmological models approximate 
reality: many independent and tight constraints on cosmological-model 
parameters allow for consistency checks on the models (see, e.g., Maor 
et al. 2002; Wasserman 2002).

For example, there is now much more higher-quality Type Ia supernova 
redshift-magnitude data. Recent applications of the redshift-magnitude 
test based on this data (see, e.g., Riess et al. 1998; Perlmutter et al. 
1999; Podariu \& Ratra 2000; Waga \& Frieman 2000; Leibundgut 2001) 
indicate that the energy density of the current universe is dominated 
by a cosmological constant, $\Lambda$, or by a term in the stress-energy 
tensor that only varies slowly with time and space and so behaves like 
$\Lambda$.\footnote{ 
See, e.g., Peebles (1984), Ratra \& Peebles (1988), Efstathiou, Sutherland, 
\& Maddox (1990), Ratra et al. (1997, 1999), Steinhardt (1999), 
Sahni \& Starobinsky (2000), Brax, Martin, \& Riazuelo (2000), and Carroll (2001) for discussions of such models.} 
Supporting evidence for $\Lambda$ or a $\Lambda$-like term is provided 
by a combination of low dynamical estimates for the non-relativistic 
matter density parameter $\Omega_0$ (see, e.g., Peebles 1993) and 
evidence for a vanishing curvature of spatial hypersurfaces from cosmic 
microwave background anisotropy measurements (see, e.g., Podariu et al. 
2001; Baccigalupi et al. 2002; Scott et al. 2002; Mason et al. 2002).

Evidence against the large value of the cosmological constant density 
parameter $\omegal$ favored by the above tests comes from estimates of 
the observed rate of multiple images of radio sources or quasars, 
produced by gravitational lensing by foreground galaxies (see, e.g., 
Ratra \& Quillen 1992; Helbig et al. 1999; Waga \& Frieman 2000; 
Ng \& Wiltshire 2001).

An improvement in data quality, as well as data from other cosmological 
tests, will be needed to resolve this situation. In the near future the 
redshift-counts test appears to be promising (see, e.g., Newman \& Davis 
2000; Huterer \& Turner 2001; Podariu \& Ratra 2001; Levine, Schulz, \& 
White 2002).

Present redshift-angular size data provide a useful consistency check. 
The redshift-angular size relation is measured by Buchalter et al. (1998) 
for quasars, by Gurvits, Kellermann, \& Frey (1999) for compact radio 
sources, and by  Daly \& Guerra (2002) for FRIIb radio galaxies. 
Vishwakarma (2001), Lima \& Alcaniz (2002), and Chen \& Ratra (2003) 
use the Gurvits et al. (1999) data to set constraints on cosmological 
parameters. Guerra, Daly, \& Wan (2000), Guerra \& Daly (1998),  
Daly, Mory, \& Guerra (2002), and Daly \& Guerra (2002) 
examine FRIIb radio galaxy redshift-angular 
size cosmological constraints in various models using the modified
standard yardstick method proposed by Daly (1994).  

Here we use the FRIIb radio galaxy redshift-angular size data of 
Guerra et al. (2000) to derive constraints on the parameters of 
a spatially-flat model with a dark energy scalar field ($\phi$) with 
scalar field potential energy density $V(\phi)$ that at low redshift 
is $\propto \phi^{-\alpha}$, $\alpha > 0$ (Peebles \& Ratra 1988). 
The energy density of such a scalar field 
decreases with time, behaving like a time-variable $\Lambda$.

We adopt the analysis technique of Guerra et al. (2000), marginalizing 
over their parameter $\beta$ to account for the uncertainty in the 
linear size of the ``standard rod" used in the redshift-angular size test, 
to derive the likelihood (probability distribution) of the scalar 
field model parameters, $L(\Omega_0, \alpha)$. This likelihood function 
is used to determine confidence regions for the model parameters. 
We compute $L(\Omega_0, \alpha)$ over the ranges $ 0.05 \leq \Omega_0 
\leq 0.95$ and $ 0 \leq \alpha \leq 8$. The radio galaxy 3C 427.1 (of 
the twenty used in the analysis) is a disproportionate contributor to 
$\chi^2$, so we present results both including and excluding this radio 
galaxy. When we exclude 3C 427.1 we renormalize the error bars to make 
the best-fit reduced $\chi^2$ unity.

\section{Results and Discussion} 

Figures 1 and 2 show the constraints on $\Omega_0$ and $\alpha$ in 
the dark energy scalar field model with $V(\phi) \propto \phi^{-\alpha}$, 
including and excluding 3C 427.1. In both cases the constraints shown 
here are consistent with, but tighter than, those derived using the 
Gurvits et al. (1999) compact radio source redshift-angular size data 
(Chen \& Ratra 2003, Fig. 3). They are also consistent with, but not as 
constraining as, those derived from the Riess et al. (1998) and Perlmutter 
et al. (1999) Type Ia supernova redshift-magnitude data (Podariu \& Ratra 
2000; Waga \& Frieman 2000). Consistent with these analyses, the analysis 
here also does not rule out large values of $\alpha$ when $\Omega_0$ is
small.

The radio galaxy 3C 427.1 can easily be identified as an outlier.  
This can be seen in Table 2 and Figures 2b, 8b, and 8c of 
Guerra \& Daly (1998), and the effective Hubble diagrams
shown in Figure 7 of Guerra et al. (2000) and
Figure 8 of Daly \& Guerra (2001), for example.  In 
the fits presented in each of these papers and those
presented by Daly \& Guerra (2002), as well
as in the fits presented here, this one source
contributes about one-half of the total $\chi^2$,
and the total $\chi^2$ for the best fit parameters is always
about 16 (see Figure 1 of Daly \& Guerra 2002, for example).
Each fit has 16 degrees of freedom since there
are 20 radio galaxy points, the cosmological model has
2 free parameters, and the radio galaxy modified
standard yardstick model has 2 free parameters.  

To date, this point has been included in all of the fits.
It is not clear whether it should be included in the fits,
or whether it should be flagged as an outlier and removed
from the data set, as was done for supernovae Type Ia by
Perlmutter et al. (1999).  If it is flagged as an outlier
and removed from the data set, the low reduced $\chi^2$
of about one-half that would result for the best fitting
parameters in any of the cosmological models considered
would indicate that the error bars on 
each radio galaxy data point have been over-estimated by a
factor of about 1.4.  It is certainly possible for these
error bars to have been overestimated; the determination
of the error bars is discussed in detail in the Appendix
of Guerra et al. (2000).  This decrease in the error bar
per point tightens the constraints on all parameters, 
including cosmological parameters and radio galaxy model
parameters, as can be seen by comparing Figures 1 and 2.  
However, the only empirical basis for removing 3C 427.1 from the
data set is it's position on the radio galaxy effective
Hubble diagram.   It's radio structure is not unusual or
remarkable (see Leahy, Muxlow, \& Stephens 1989).  The use
of this radio source to determine the ambient gas density
(Wellman et al. 1997a), the ambient gas temperature 
(Wellman et al. 1998b), and the synchrotron aging independent
ambient gas pressure, the beam power, and the total
AGN-jet lifetime of this source (Wan, Guerra, \& Daly 2000)
are all unremarkable and
quite in line with sources with similar redshifts, sizes,
and radio powers.  The determination of the source
redshift seems secure (Spinrad, Stauffer, \& Butcher 1985).
It does not appear to be an especially bright infrared source;
upper bounds on its infrared luminosity are presented by 
Meisenheimer et al. (2001).  Thus, it is unlikely to be an
especially highly obscured quasar that should have been removed from 
this sample of radio galaxies.   And, it's radio morphology and environmental
properties are not unusual (Harvanek \& Stocke 2002).  Thus,
there does not seem to be any empirical basis to remove it
from the radio galaxy sample other than the small 
value of its predicted average size 
$D_*$ relative to the average size of the full population of
FRIIb radio galaxies $\langle D \rangle$ at similar redshifts.    

Additional radio data could substantially improve the constraints
presented here on the dark energy scalar field model, and on other
models.  Of the 70 sources in the 
parent population, 20 have values of $D_*$ determined.  The radio 
data for these 20 radio sources were available in the 
published literature and in the VLA archive, and were originally obtained 
for studies other than those presented here.  Ten additional 
sources will be observed at the VLA this year; these data are optimized to
deterime cosmological parameters and to test and constrain the
underlying radio galaxy model.  The 40 remaining sources will
then be observed, which will improve the radio galaxy constraints
enormously.  The improvement will come from the additional
number of data points and, with the very high quality data
expected, the error bar per point may be smaller than that
obtained using published data.   

It is encouraging that the FRIIb radio galaxy redshift-angular size data 
constraints are consistent with and not much weaker than those derived 
from Type Ia supernova redshift-magnitude data. Future higher-quality 
redshift-angular size data is eagerly anticipated.

\bigskip

We acknowledge helpful discussions with Joel Carvalho,
Megan Donahue, Eddie Guerra, Philip Mannheim, Chris O'Dea, 
Adam Reiss, and the referee Dave Helfand.  We 
acknowledge support from NSF CAREER grant AST-9875031, NSF NYI grant 
AST-0096077, NSF grant AST-0206002, and Penn State University.


\begin{figure}[p] 
\psfig{file=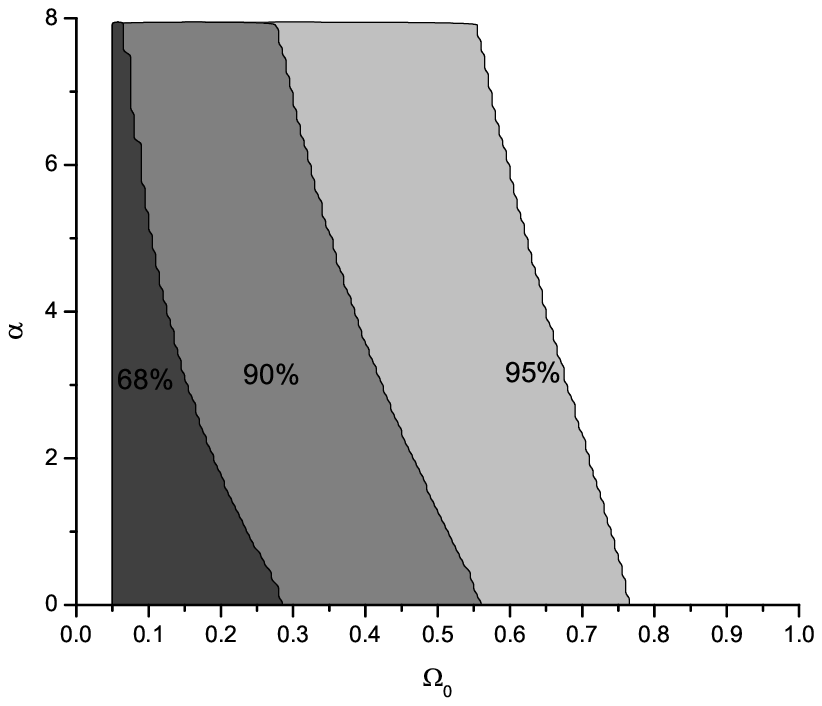,height=7.0in,width=6.7in,angle=0} 
\caption{Confidence contours for the dark energy scalar field model 
with inverse power-law potential energy density $V(\phi) \propto 
\phi^{-\alpha}$, derived using all twenty radio galaxies (i.e., 
including 3C 427.1).} 
\end{figure}
\begin{figure}[p] 
\psfig{file=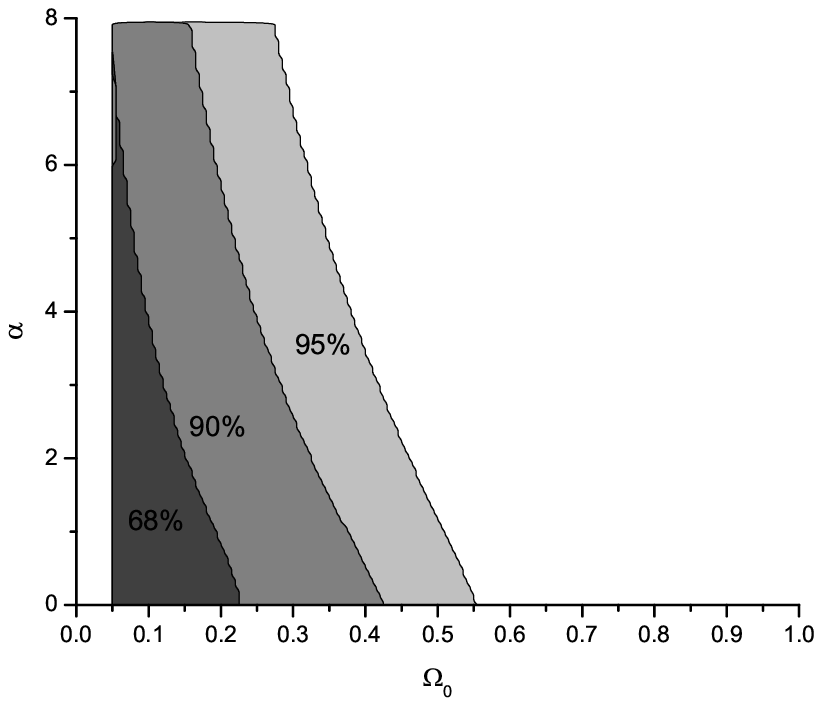,height=7.0in,width=6.7in,angle=0} 
\caption{Confidence contours for the dark energy scalar field model 
with inverse power-law potential energy density $V(\phi) \propto 
\phi^{-\alpha}$, derived using only nineteen radio galaxies (i.e., 
excluding 3C 427.1).} 
\end{figure}
\end{document}